\documentclass[sigconf]{acmart}

\usepackage{amsmath}
\usepackage{booktabs}
\usepackage{placeins}

\renewcommand\footnotetextcopyrightpermission[1]{}
\graphicspath{{figures/}}

\newcommand{\clip}{\operatorname{clip}_{[0,1]}}
\newcommand{\targetwl}{1310}

\AtBeginDocument{%
}

\title{When Every Simulation Counts: Value-Based Reinforcement Learning for Accelerated Photonics Inverse Design}

\author{Longying Wen}
\authornote{Longying Wen, Feiyang Wu, and Jinglin Yu contributed equally to this work.}
\affiliation{%
  \department{School of Science and Engineering}
  \institution{The Chinese University of Hong Kong, Shenzhen}
  \city{Shenzhen}
  \country{China}
}
\email{225015041@link.cuhk.edu.cn}

\author{Feiyang Wu}
\authornotemark[1]
\affiliation{%
  \department{School of Science and Engineering}
  \institution{The Chinese University of Hong Kong, Shenzhen}
  \city{Shenzhen}
  \country{China}
}
\email{fywu2003@gmail.com}

\author{Jinglin Yu}
\authornotemark[1]
\affiliation{%
  \department{School of Science and Engineering}
  \institution{The Chinese University of Hong Kong, Shenzhen}
  \city{Shenzhen}
  \country{China}
}
\email{121090729@link.cuhk.edu.cn}

\author{Chongxian Yuan}
\affiliation{%
  \department{School of Science and Engineering}
  \institution{The Chinese University of Hong Kong, Shenzhen}
  \city{Shenzhen}
  \country{China}
}
\email{cxyuan@cuhk.edu.cn}

\author{Renjie Li}
\authornote{Corresponding authors: Zhaoyu Zhang (primary) and Renjie Li (secondary).}
\affiliation{%
  \department{Nick Holonyak Micro and Nanotechnology Laboratory}
  \institution{University of Illinois Urbana-Champaign}
  \city{Urbana}
  \state{Illinois}
  \country{USA}
}
\email{renjie2@illinois.edu}

\author{Zhaoyu Zhang}
\authornotemark[2]
\affiliation{%
  \department{School of Science and Engineering}
  \institution{The Chinese University of Hong Kong, Shenzhen}
  \city{Shenzhen}
  \country{China}
}
\email{zhangzy@cuhk.edu.cn}

\begin{abstract}
Photonic-crystal surface-emitting lasers (PCSELs) can combine high-power operation with narrow-divergence surface emission, but optimizing coupled parameters requires costly full-wave simulations. Deep Q-network (DQN) optimization can reuse simulated transitions to guide edits, yet which value-learning mechanisms remain reliable under tight simulation budgets is unknown. We address this gap by comparing baseline DQN and six value-based variants for a seven-variable PCSEL design under a shared objective, simulator, 83-call budget, and four matched initializations. Beyond endpoints, we analyze sample efficiency, policy behavior, and physical response to separate learning gains from favorable starts or exploratory jumps. Dueling DQN is the only variant to improve all four seeds. Relative to the first evaluated designs, its selected structures increase the mean quality factor ($Q$) from $2.18\times10^5$ to $5.63\times10^6$ ($25.8\times$), reduce wavelength error by 64\%, and increase upward power by 47\%; compared with baseline DQN, they achieve a $2.6\times$ higher mean $Q$ under the same budget. Other variants yield no consistent improvement; Double DQN reproduces baseline trajectories, while Rainbow-lite shows high upside but strong seed dependence. These results identify Dueling DQN as the most reliable configuration tested for simulation-budget-limited PCSEL inverse design and provide a reproducible framework for attributing algorithmic gains in scientific optimization.
\end{abstract}



\begin{document}
\maketitle

\begingroup
\small
\noindent\raggedright\textbf{Resource Availability:}\\
The source code and experiment scripts used in this study are publicly available at
\url{https://github.com/Longying-Wen/PCSEL-RL}.
\par
\endgroup

\section{Introduction}

High-power semiconductor lasers typically face a trade-off between output power and beam quality. Broad-area semiconductor lasers can supply high power but tend toward multimode emission, whereas vertical-cavity surface-emitting lasers (VCSELs) provide surface-normal beams but are difficult to scale to high single-mode power \cite{yoshida2023highbrightness,hirose2014watt}. Photonic-crystal surface-emitting lasers (PCSELs) use a two-dimensional photonic crystal to couple a large emitting area into a coherent mode, enabling high-power, narrow-divergence surface emission \cite{noda2017pcsel,ishizaki2019progress}. This combination makes PCSELs promising compact sources for material processing, sensing, and free-space optics. Each candidate nevertheless requires a full-wave finite-difference time-domain (FDTD) solution of Maxwell's equations \cite{yee1966numerical,taflove2005computational}. Such evaluations are computationally expensive \cite{schneider2019benchmark}, limiting the number of designs that can be tested in practice.

Realizing this combination of power, beam quality, and spectral control is, however, a demanding inverse-design problem. Layer thicknesses, refractive indices, lattice geometry, and air-hole dimensions jointly determine wavelength, quality factor ($Q$), vertical extraction, mode area, and far-field divergence. These responses are strongly coupled: a high-$Q$ design may still be unsuitable when its resonance is off target or its upward flux is weak, whereas stronger extraction can shorten photon lifetime \cite{liang2011coupled}. The resulting nonconvex, multi-objective landscape must therefore be explored within a small budget of costly full-wave evaluations.

Several inverse-design strategies address this cost in different ways. Black-box methods target expensive objectives \cite{jones1998efficient,snoek2012practical,shahriari2016taking}. Topology optimization spans many geometric degrees of freedom \cite{jensen2011topology,molesky2018outlook,christiansen2021tutorial}, whereas adjoint and automatic-differentiation methods provide efficient gradients when the solver permits \cite{hughes2018adjoint,minkov2020autodiff}. Fabrication-constrained demonstrations \cite{piggott2015inverse,piggott2017fabrication} and reusable frameworks \cite{su2020spins} improve practicality. Neural surrogates and generative models reduce online solves only after a simulation dataset is assembled \cite{peurifoy2018nanophotonic,liu2018generative,asano2019iterative}. Reviews synthesize their scope and limitations \cite{ma2021deep,wiecha2021deep,jiang2021deep}, while direct benchmarks expose data and nonuniqueness challenges \cite{ma2022benchmarking}.

Reinforcement learning (RL) offers a complementary route for allocating simulator calls adaptively, without requiring a fixed inverse-design database before optimization begins \cite{sutton2018reinforcement}. An agent can acquire experience while modifying the device, receive the full-wave response of each candidate, and condition later edits on the trajectory already observed. This sequential view avoids solving a one-shot mapping from a desired spectrum to a unique geometry and instead asks which nearby structural change is worth the next expensive query. Prior work has demonstrated this potential in thin films, dielectric nanostructures, hybrid supervised--RL workflows, and photonic-crystal resonators \cite{jiang2020thinfilm,yeung2022hybrid,li2023drl}. PCSEL design has also been studied through rigorous threshold modeling, quantum annealing, and sequence-modeling formulations \cite{song2018threshold,inoue2022quantum,zhang2024sequence}, reflecting a broader move from manual iteration toward automated scientific design.

The unresolved issue is not whether an RL loop can produce one improved structure, but which learning mechanisms remain dependable when the entire experiment contains only tens of full-wave calls. A high endpoint may reflect a consistently useful policy, a favorable random initialization, an isolated exploratory jump, or early commitment to a fortuitous basin. These explanations are difficult to separate when methods use different warm starts, query budgets, or reporting conventions. We therefore treat the simulator call as the common unit of cost and compare matched trajectories rather than only final designs.

RL is valuable in this regime not only because it collects data online, but because a shared value function can reuse each transition to inform later nearby edits without evaluating every neighbor. The discrete policy also leaves an interpretable path through thickness, index, hole-size, and lattice changes that can be audited against the physical outputs. Yet enhancements developed for large-data control do not necessarily transfer to a replay buffer containing only dozens of post-initialization samples. Dueling heads, return distributions, prioritized replay, parameter noise, and Double targets address different bottlenecks while adding their own estimation assumptions. Their relevance must therefore be established under sparse, coupled physical feedback with matched starts, budgets, and process diagnostics.

We formulate PCSEL inverse design as sequential edits of seven active variables and compare DQN with dueling, C51 distributional, prioritized-replay (PER), Noisy, Double, and Rainbow-lite variants. All configurations share four seeds, an 83-call budget, the same simulator and local-search schedule, and a fixed multi-physics score. We report endpoint, sample-efficiency, policy-behavior, and physical-response metrics. Dueling DQN is the only variant that improves all seeds and moves the selected devices into a higher-$Q$ regime; Double DQN reproduces the baseline trajectories, while Rainbow-lite reveals high upside but strong seed dependence. The contribution is therefore both a controlled PCSEL benchmark and an empirical account of which value-learning mechanisms remain reliable under a scarce simulation budget.

\begin{figure}[!ht]
\centering
\includegraphics[width=\columnwidth]{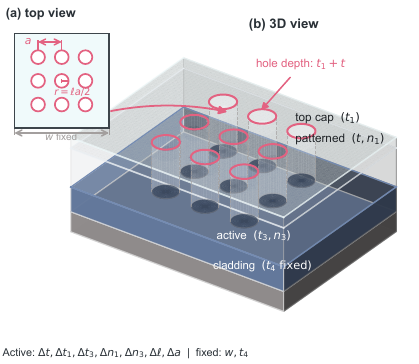}
\Description{Compact schematic of the PCSEL template. A small inset shows the top view of a three-by-three air-hole lattice, with lattice constant a, radius r equals ell a over two, and fixed lateral width w. A curved arrow links the inset to the transparent three-dimensional multilayer cavity, which shows the top cap, patterned layer, active layer, fixed lower cladding, etched holes, and seven active structural controls.}
\caption{PCSEL geometry and controls. (a) Inset top view defines $a$, $r=\ell a/2$, and fixed $w$. (b) Transparent multilayer view locates the seven optimized variables; the holes have depth $t_1+t$, while $w$ and $t_4$ remain fixed.}
\label{fig:pcsel_structure}
\end{figure}

\section{Design Task and Objective}

\subsection{PCSEL Template and FDTD Evaluation}

Each candidate is a seven-dimensional offset vector
\begin{equation}
    \mathbf{s} =
    [\Delta t,\Delta t_1,\Delta t_3,\Delta n_1,\Delta n_3,\Delta \ell,\Delta a]^\mathsf{T}.
\end{equation}
Here $t,t_1,t_3$ are layer thicknesses, $n_1,n_3$ are refractive indices, $\ell$ controls air-hole size, and $a$ is the lattice constant; the hole radius is $r=\ell a/2$. Lateral width $w$ and lower-cladding thickness $t_4$ remain fixed. The environment converts $\mathbf{s}$ to an absolute geometry, executes the Lumerical FDTD solver, and returns one physical response per costly query. Figure~\ref{fig:pcsel_structure} locates all seven controls.

The offset representation preserves a common nominal cavity while allowing every action to be interpreted as a manufacturable geometric or material edit. In the top view, $a$ controls the center-to-center spacing of the representative $3\times3$ hole pattern and $\ell$ changes the hole radius through their product. In the multilayer view, the etched holes traverse the top cap and patterned layer with total depth $t_1+t$ and terminate at the active-layer interface. The remaining thickness and index coordinates change the vertical optical path, material contrast, and modal overlap without changing the lateral device footprint. Keeping $w$ and $t_4$ fixed therefore isolates a seven-variable search while retaining the principal in-plane and vertical controls that govern the simulated cavity response.

\subsection{Physics-Aware Score}

FDTD returns
\begin{equation}
    \mathbf{y}(\mathbf{s})=[Q,\lambda,P,A_\mathrm{eff},\theta]^\mathsf{T},
\end{equation}
where the entries are cavity quality factor, resonant wavelength, upward power ratio, effective mode area, and divergence angle. With target wavelength $\targetwl$ nm, the normalized components and fixed scalar objective are
\begin{align}
    c_{Q,\mathrm{raw}} &= \clip\left(\frac{\log_{10}(\max(Q,10^3))-3.5}{3.0}\right), \\
    c_{\lambda} &= \exp\left[-\left(\frac{\lambda-\targetwl}{5}\right)^2\right], \\
    c_Q &= c_{Q,\mathrm{raw}}\left(0.7+0.3c_{\lambda}\right), \\
    c_P &= \clip\left(\frac{P}{0.15}\right), \\
    c_A &= \clip\left(\frac{A_\mathrm{eff}}{1.2\times 10^{-13}}\right), \\
    c_\theta &= \clip\left(1-\frac{\max(\theta-1,0)}{0.5}\right), \\
    S &= 60c_Q + 18c_{\lambda} + 8c_P + 4c_A \notag\\
      &\quad + 10c_{\theta} - 10\mathbf{1}_\mathrm{terminal}.
    \label{eq:score}
\end{align}

Equation~\ref{eq:score} encodes a hierarchy of engineering preferences rather than treating the outputs as commensurate physical units. Log scaling accommodates the orders-of-magnitude range of cavity lifetimes and gives $Q$ the largest share. Spectral alignment enters directly through $c_\lambda$ and also gates $c_{Q,\mathrm{raw}}$, so an off-target high-$Q$ mode cannot receive the full confinement reward. The upward-power term favors useful surface extraction, the area term rewards an extended cavity mode, and $c_\theta$ penalizes excessive far-field divergence. Clipping introduces diminishing returns once a secondary requirement is satisfied and prevents one auxiliary quantity from compensating arbitrarily for poor resonance quality or wavelength mismatch. Thus, $S$ is a fixed search utility for ranking simulated candidates, not a new physical observable or a universal PCSEL merit function. We keep it unchanged across all algorithm groups and report the raw $Q$, wavelength, power, area, and divergence alongside score-based comparisons.

\subsection{Sequential Local-Edit Formulation}

The optimizer changes the device through local edits rather than proposing an unrelated geometry at every step. The action space contains paired increases and decreases for each active coordinate: 2.5 nm for thickness and lattice-constant offsets and 0.005 for refractive-index and hole-factor offsets. These steps are small enough to trace the response of a nearby physical design but large enough to produce resolvable changes in the full-wave output. The same discrete definition is used for every algorithm, so differences in query allocation cannot be attributed to a different geometric proposal mechanism.

At call $k$, the agent observes $\mathbf{s}_k$, selects $a_k$, receives $\mathbf{s}_{k+1}$, and obtains reward $r_k=S(\mathbf{s}_{k+1})$. A trajectory is consequently an auditable sequence of parameter edits connected to absolute FDTD geometries and five raw physical responses. This trace connects score improvements to stored device structures and supports the structure--response and consistency analyses below.

\section{DQN Methods and Experimental Setup}
\label{sec:method}

\subsection{Baseline DQN}

The discrete local-edit space and the cost of each FDTD evaluation motivate a value-based formulation. At state $\mathbf{s}$, DQN estimates the values of all 14 active edits without evaluating every neighboring geometry. Each simulated transition $(\mathbf{s},a,r,\mathbf{s}')$ enters replay and can therefore support multiple network updates. Following Q-learning and DQN \cite{watkins1992qlearning,mnih2015human}, the bootstrap target is
\begin{equation}
    y_\mathrm{DQN}=r+\gamma \max_{a'}Q_{\theta^-}(\mathbf{s}',a'),
\end{equation}
where $\theta^-$ denotes the target-network parameters. Replay reuses scarce physical evaluations, while the target network limits feedback from rapidly changing value estimates. DQN provides the common reference because the tested extensions modify identifiable parts of this estimator---the value head, return representation, replay distribution, exploration rule, or bootstrap target---without changing the PCSEL task or action space.

\subsection{DQN Variants}

Dueling DQN separates the value of the current design neighborhood from the relative advantage of an individual edit \cite{wang2016dueling}:
\begin{equation}
    Q(\mathbf{s},a)=V(\mathbf{s})+
    \left(A(\mathbf{s},a)-\frac{1}{16}\sum_{j=0}^{15}A(\mathbf{s},j)\right).
\end{equation}
The compatibility head is mean-centered before the two fixed-width outputs are masked. This decomposition tests whether recognizing a promising local neighborhood is easier than immediately ranking 14 edits with similar and coupled physical effects.

Distributional DQN uses C51 to predict a categorical return distribution over 51 fixed atoms and selects actions using its expectation \cite{bellemare2017distributional},
\begin{equation}
    Q(\mathbf{s},a)=\sum_i z_i p_\theta(z_i\mid\mathbf{s},a).
\end{equation}
It tests whether sharp score changes are represented more effectively by a return distribution, against the sample demand of estimating 51 probabilities from a short trajectory. PER retains the scalar value head but samples transitions according to temporal-difference error \cite{schaul2016prioritized}; it tests whether rare responses merit repeated updates, while recognizing that large errors can also arise from unproductive spectral detuning.

Noisy DQN replaces post-initialization epsilon exploration with learned parameter noise in the later hidden layers and output head \cite{fortunato2018noisy}, thereby testing adaptive parameter-space exploration. Double DQN selects the next action with the online network and evaluates it with the target network \cite{vanhasselt2016double}, isolating whether maximization bias changes the realized search. Rainbow-lite combines Double targets, dueling heads, PER, C51, and NoisyLinear exploration while retaining one-step transitions \cite{hessel2018rainbow}. It examines interactions among these mechanisms rather than reproducing every component of the original Rainbow agent.

\subsection{Experimental Protocol}

\paragraph*{Matched runs.}
Every run uses the same PCSEL template, objective, seeds 0--3, and budget of 83 FDTD calls. The first 20 calls are shared random-start evaluations within each seed. Learning begins after call 20, and the remaining 63 evaluations form three-edit windows anchored at the current incumbent. Algorithms compared within a seed therefore receive the same initialization and differ only in post-initialization query allocation; the four seeds act as matched experimental blocks. Initialization is an experimental control, not a claimed improvement mechanism.

The implementation is held fixed wherever the tested mechanism permits. Scalar variants share an 80--120--80 multilayer perceptron, replay capacity, optimizer, discount, target-update interval, and training frequency. A 16-output compatibility head is retained, but the two fixed-width actions are masked, leaving the 14 active edits in Section~2. Dueling and distributional variants replace only the corresponding output structure, Noisy DQN changes the exploration layers, and PER changes replay sampling. Full hyperparameters are provided in Appendix~\ref{app:implementation}.

\paragraph*{Evaluation measures.}
The primary endpoint is the best score reached within 83 calls, summarized by its cross-seed mean, minimum, standard deviation, and paired change from DQN. Sample efficiency is assessed using the discrete mean of the best-so-far score over calls 21--83 (AUC) and the first call reaching $S\geq83$, averaged over successful crossings and reported with the success count. Candidate quality is characterized by the post-initialization Top-5 mean and the number of seeds reaching $S\geq90$. Post-initialization, late-stage, and policy-only summaries distinguish persistent improvement from a single extreme; policy-selected $S\geq83$ calls count useful non-random selections, or noisy-greedy selections for Noisy DQN and Rainbow-lite. The five raw FDTD outputs of each objective-selected device are also reported so that a scalar improvement is not treated as a sufficient physical description. This seed-aware, multi-metric assessment follows recommended practice for small-sample RL comparisons \cite{henderson2018matters,agarwal2021statistical}.

\paragraph*{Structure--response statistics.}
For the local diagnostic, calls 21--83 are converted to increasing-parameter finite differences. Reverse traversals are reoriented to the same positive-edit convention, and repeated visits to an identical geometry edge are collapsed before medians, quartiles, and improving-edge fractions are computed. The statistics therefore describe unique policy-sampled transitions rather than weighting repeated exploitation as independent evidence. All 28 matched logs are complete; their 1,764 valid post-initialization transitions collapse to 903 unique local edges. Because no counterfactual neighbors are simulated, these quantities are interpreted as policy-conditioned local associations rather than global causal sensitivities. Logs retain every state, action, score, physical response, exploration flag, and incumbent; objective-selected configurations are saved as complete simulation projects for the consistency checks reported in Section~4.1 and Appendix~\ref{app:endpoint_audit}.

\section{Results and Discussion}

\subsection{Optimization Results}

\begin{figure}[t]
\centering
\includegraphics[width=\columnwidth]{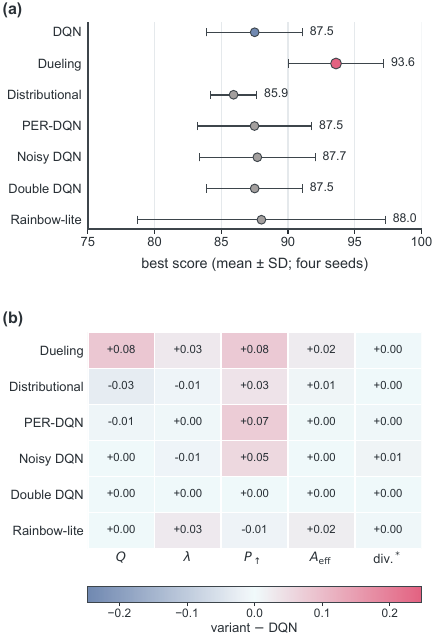}
\Description{Two vertically stacked panels formatted for a single column. Panel a is a horizontal point-and-error plot of mean best score for DQN and six matched variants, with one-standard-deviation error bars across four seeds. Panel b is a heatmap of cross-seed mean normalized physical-component differences between each variant and its matched DQN baseline; muted blue denotes negative differences and rose denotes positive differences.}
\caption{Matched endpoint and physical comparison. (a) Mean best score $\pm$ one standard deviation over four seeds. (b) Mean normalized component change relative to DQN; positive is better. \textit{div.*} is a stricter visualization-only scale. Seed-level panels are in Appendix Figure~\ref{fig:component_delta_by_seed}.}
\label{fig:matched_overview}
\label{fig:component_delta}
\end{figure}

Figure~\ref{fig:matched_overview}(a) summarizes the endpoint ranking. Dueling DQN raises the mean best score from 87.51 to 93.60, improves all four matched seeds, and raises the minimum from 84.65 to 88.70 without increasing cross-seed dispersion. Because each comparison inherits the same seed-specific initialization incumbent, these four gains reflect differences after initialization rather than unrelated starting designs. Seed-level endpoints and first-best calls are reported in Appendix Table~\ref{tab:dqn_variants}.

\begin{figure}[t]
\centering
\includegraphics[width=\columnwidth]{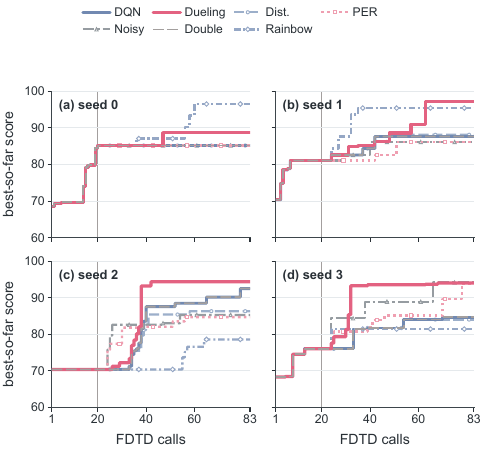}
\Description{Four best-so-far learning curves, one for each seed, comparing seven matched DQN configurations under an 83-call budget. Color is paired with distinct line styles and markers. Double DQN is shown as a thin dashed line exactly overlapping DQN.}
\caption{Best-so-far score by seed. The vertical line ends the 20-call initialization; dashed Double DQN overlaps DQN.}
\label{fig:dqn_variant_curves}
\end{figure}

Figure~\ref{fig:dqn_variant_curves} resolves the seed dependence hidden by the mean. Distributional DQN never reaches 90; PER-DQN and Noisy DQN remain near DQN on average but each produces one strong seed3 run; and Double DQN reproduces every DQN action and physical trajectory. Rainbow-lite reaches 96.54 and 95.43 in seeds0--1 but only 78.62 and 81.46 in seeds2--3, giving the largest standard deviation. Dueling DQN is therefore the only tested modification that shifts all four paired outcomes rather than producing isolated gains.

The endpoint differences correspond to physically distinct devices. Figure~\ref{fig:matched_overview}(b) shows positive mean changes for dueling in normalized $Q$, wavelength alignment, upward power, and mode area. Two selected endpoints clarify why the scalar ranking is not determined by $Q$ alone. The highest-scoring design reaches 97.283 with $Q=4.533\times10^6$, $\lambda=1310.132$~nm, and upward power 0.124. The highest-$Q$ dueling endpoint reaches $1.2015\times10^7$ at the same wavelength but scores 94.045 with upward power 0.062.

\begin{figure}[!t]
\centering
\includegraphics[width=\columnwidth]{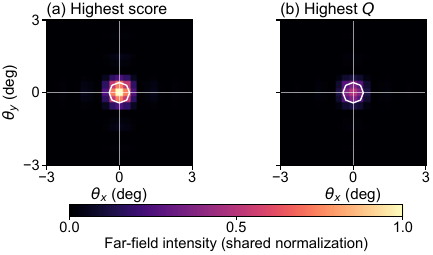}
\Description{Shared-scale far-field intensity maps of the highest-scoring and highest-Q Dueling DQN endpoints. Both have an on-axis lobe, while the highest-scoring design has the greater intensity.}
\caption{Shared-scale far-field intensity of (a) the highest-scoring and (b) the highest-$Q$ Dueling DQN endpoints. Intensities are normalized to the peak of panel (a); white contours mark half of each panel's local maximum. Full near- and far-field profiles and monitor details are given in Appendix Figure~\ref{fig:representative_fields}.}
\label{fig:representative_farfields_main}
\end{figure}

The shared-scale projections in Figure~\ref{fig:representative_farfields_main} provide a direct view of this tradeoff. Both endpoints retain an on-axis lobe with the same reported $0.841^\circ$ divergence, but the peak and integrated intensities of the highest-scoring design are $2.004\times$ those of the highest-$Q$ design, consistent with the $2.007\times$ ratio of their upward powers. Thus, the 2.65-fold increase in $Q$ neither narrows the projected beam nor compensates for the reduction in vertical extraction.

For both designs, $Q>10^{6.5}$, so $c_{Q,\mathrm{raw}}$ in Equation~\ref{eq:score} is clipped at one; their identical wavelengths also give the same $Q$-gating and spectral contributions. The power terms contribute 6.626 and 3.302 points, whereas the highest-$Q$ design recovers only 0.086 points through its slightly larger mode area, leaving the observed 3.238-point gap. Once resonance quality and alignment satisfy the clipped objective, vertical extraction therefore determines this ranking; the lower-$Q$ endpoint is not asserted to be universally superior. Independent re-simulation of the saved geometries reproduces the recorded responses and scores to the reported precision (Appendix Table~\ref{tab:physical_metrics}).

\subsection{Performance of Dueling DQN}

\begin{table*}[!t]
\centering
\caption{Matched multi-metric comparison. Higher is better except for standard deviation and wavelength error. AUC covers calls 21--83; Top-5 averages the strongest post-initialization candidates; policy $\geq83$ counts non-random selections. Physical columns average each seed's objective-selected device.}
\label{tab:multimetric_comparison}
\resizebox{\textwidth}{!}{\begin{tabular}{lrrrrrrrrrr}
\toprule
& \multicolumn{3}{c}{Endpoint} & \multicolumn{2}{c}{Efficiency} & \multicolumn{2}{c}{Robustness} & \multicolumn{3}{c}{Physical response} \\
\cmidrule(lr){2-4}\cmidrule(lr){5-6}\cmidrule(lr){7-8}\cmidrule(lr){9-11}
Model & Mean best & Min. & Std. & AUC & Top-5 & $\geq$90 seeds & Policy $\geq$83 & $Q$ ($10^6$) & $|\Delta\lambda|$ & $P$ \\
\midrule
DQN & 87.51 & 84.65 & 3.59 & 84.35 & 85.77 & 1/4 & 3.00 & 2.19 & 0.86 & 0.078 \\
\textbf{Dueling} & \textbf{93.60} & \textbf{88.70} & 3.58 & \textbf{89.22} & \textbf{93.12} & \textbf{3/4} & 7.25 & \textbf{5.63} & \textbf{0.26} & \textbf{0.089} \\
Distributional & 85.92 & 84.04 & \textbf{1.74} & 83.76 & 84.92 & 0/4 & 5.00 & 1.32 & 0.95 & 0.083 \\
PER & 87.49 & 84.72 & 4.30 & 84.08 & 85.25 & 1/4 & 1.00 & 2.61 & 0.93 & \textbf{0.089} \\
Noisy & 87.70 & 85.20 & 4.34 & 85.52 & 85.16 & 1/4 & 5.50 & 1.83 & 0.98 & 0.086 \\
Double & 87.51 & 84.65 & 3.59 & 84.35 & 85.77 & 1/4 & 3.00 & 2.19 & 0.86 & 0.078 \\
Rainbow-lite & 88.01 & 78.62 & 9.29 & 84.78 & 87.22 & 2/4 & \textbf{15.75} & 2.40 & 0.35 & 0.075 \\
\bottomrule
\end{tabular}
}
\end{table*}

Endpoint score alone would not establish a more reliable search. Under Dueling DQN, however, the weakest seed improves from 84.65 to 88.70 while the cross-seed standard deviation remains essentially unchanged (3.58 versus 3.59). The post-initialization best-so-far AUC rises from 84.35 to 89.22, the Top-5 mean from 85.77 to 93.12, and, among successful crossings, the first $S\geq83$ occurs 7.4 calls earlier on average. Three seeds reach 90, compared with one for DQN, while the mean number of post-initialization score-$\geq90$ evaluations increases from 0.50 to 12.50. The endpoint gain is therefore accompanied by earlier and repeated discovery of high-quality candidates rather than a single terminal maximum.

The shift persists across the realized query stream. The post-initialization mean rises from 74.88 to 77.40 and the calls-64--83 mean from 76.79 to 79.69, confirming that the increased yield persists late in the budget. Post-initialization score-$\geq83$ evaluations increase from 8.75 to 21.00 per seed, and policy-selected score-$\geq83$ calls increase from 3.00 to 7.25. Useful candidates are therefore not confined to epsilon-random exploration, although the strongest seed~1 endpoint is reached through two random thickness edits. Complete process statistics are reported in Appendix Table~\ref{tab:extended_benchmark_matrix}.

\begin{figure}[!b]
\centering
\includegraphics[width=0.90\columnwidth]{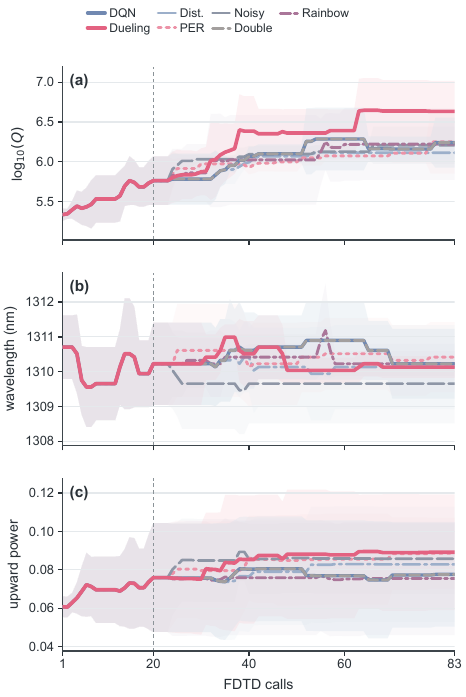}
\Description{Three vertically stacked process-level panels comparing seven matched DQN configurations over 83 FDTD calls. The panels show log Q, resonance wavelength, and upward power. Curves are cross-seed means of the physical metrics of the incumbent best design, and shaded bands indicate one standard deviation. Double DQN is dashed and overlaps DQN.}
\caption{Incumbent (a) $Q$, (b) wavelength, and (c) upward power across calls (mean $\pm$ SD). The vertical line ends initialization; dashed Double DQN overlaps DQN. Other metrics are in Appendix Figure~\ref{fig:physical_process_supplement}.}
\label{fig:dqn_dueling_physical_process}
\end{figure}

Figure~\ref{fig:dqn_dueling_physical_process} shows that this difference is also present in the raw FDTD responses of the score-selected incumbents. Relative to DQN, the objective-selected dueling structures increase mean $Q$ from $2.19\times10^6$ to $5.63\times10^6$, reduce mean wavelength error from 0.86 to 0.26~nm, and raise mean upward power from 0.078 to 0.089. At seed level, wavelength alignment improves in all four runs, while raw $Q$ and upward power each improve in three. These responses separate through multiple incumbent updates rather than only at the final call, showing that the physical gain is not a terminal outlier.

The action logs further show that no single route accounts for the gain. Seeds~0 and~3 obtain their principal improvements from policy-selected thickness edits. In seed~2, two policy-selected $+t_1$ edits relocate the incumbent before exploratory $-n_1$ and $+n_3$ probes discover the final high-score region. Seed~1 provides the complementary case: within a window anchored at 91.039, epsilon-random $+t$ and $+t_3$ edits at calls 63--64 raise the incumbent to 97.283, whereas the following greedy $+n_3$ edit scores 77.522 and is rejected. The seed~1 endpoint therefore comes from a two-step stochastic path, and those exploratory choices cannot be attributed directly to the dueling head. Across seeds, the evidence supports both direct exploitation and allocation of the remaining exploratory calls near designs with greater improvement potential.

This behavior is consistent with the information structure of local PCSEL editing. Each state offers 14 paired edits whose FDTD responses share a broad dependence on neighborhood quality but differ through weaker, coupled action effects. The value stream $V(\mathbf{s})$ can share evidence about whether the current neighborhood is promising, while $A(\mathbf{s},a)$ resolves the residual ranking among edits. The higher AUC, Top-5 quality, policy-selected high-score count, and minimum endpoint---without increased cross-seed dispersion---are consistent with this division. They do not establish the decomposition as a causal mechanism, but they show that its gain under the fixed budget is distributed across seeds, calls, and physical responses rather than being attributable to one favorable start or one random event.

\begin{figure*}[!t]
\centering
\includegraphics[width=0.94\textwidth]{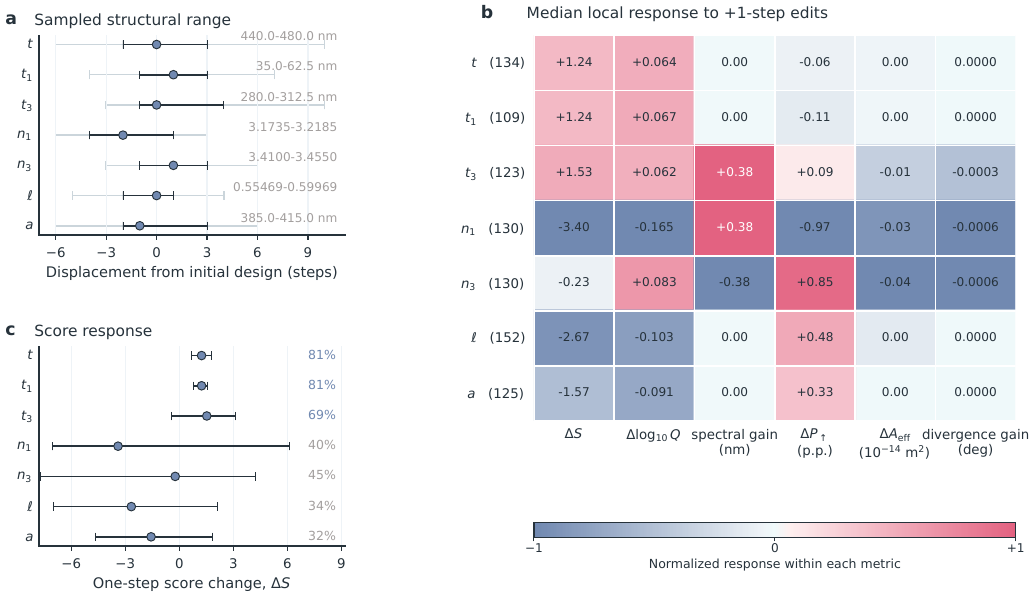}
\Description{Three-panel summary of sampled structure and local response. Panel a shows the sampled displacement range for each of the seven active variables. Panel b is a heatmap of median changes in score and five physical outputs after positive one-step edits. Panel c shows the distribution of one-step score changes and the percentage of improving sampled edges for each variable.}
\caption{Policy-sampled structure--response relations. (a) Visited displacement ranges; light whiskers denote min--max, dark whiskers 5--95\%, and points the median. (b) Raw median responses to positive one-step edits over unique geometry edges; color is normalized within each metric, and $n$ is shown in parentheses. (c) Score-response medians (points), interquartile ranges (whiskers), and improving-edge fractions (percentages). These are local associations, not global causal sensitivities.}
\label{fig:structure_action_response}
\end{figure*}

\subsection{Comparison with Other DQN Variants}

Distributional DQN does not simply fail to train. Its overall and late-stage post-initialization means (74.66 and 76.34) remain close to DQN, and it continues to generate moderate-score candidates; nevertheless, none of its four runs reaches 90. Return-distribution modeling therefore changes the realized search without improving reliable access to the high-score tail.

PER-DQN and Noisy DQN each produce one strong seed~3 trajectory. A large temporal-difference error can reflect harmful detuning or an isolated response as well as a reusable improvement direction, while parameter-space novelty need not correspond to larger $Q$ or better wavelength alignment. Both methods can exploit a favorable basin once encountered, but neither does so consistently under the present budget.

Double DQN provides the cleanest controlled null. Its target calculation and training losses differ after learning begins, yet all 83 actions, scores, and physical responses match DQN for every seed. The changed value estimates are therefore insufficient to alter an argmax under the shared random stream, although this result is specific to the present replay size, target-update interval, and local horizon.

Rainbow-lite shows why combined effects cannot be inferred from isolated components. Its post-initialization and late means rise to 77.67 and 80.40, but all score-90 evaluations remain confined to seeds~0--1. The combination can concentrate queries rapidly; whether this helps depends on the physical basin favored early in a run. Rainbow-lite is therefore interaction-sensitive rather than uniformly stronger. The corresponding structural directions are examined below.

Together, the contrast variants show that target correction, replay prioritization, return modeling, and alternative exploration can change query allocation without improving cross-seed reliability. Among the isolated modifications, only value--advantage decomposition produces a repeatable shift in matched outcomes; combining several mechanisms increases both achievable performance and sensitivity to the early trajectory.

\subsection{Structure--Response Analysis}

The endpoint and trajectory comparisons do not identify which structural directions produce the observed physical tradeoffs. Figure~\ref{fig:structure_action_response} therefore links unique policy-sampled geometry edges to their score and raw FDTD responses. Positive $t$ and $t_1$ edits improve about 81\% of sampled edges, while increasing $t_3$ gives the largest median score change (+1.53) together with positive $Q$, spectral-alignment, and power responses. Increasing $n_1$, $\ell$, or $a$ is negative on median; increasing $n_3$ raises median $Q$ and power but loses wavelength alignment, leaving the median score nearly unchanged.

These trends are physically plausible. Changes in $t$ and $t_1$ modify the optical thickness of the patterned region and top cap and therefore perturb the vertical interference condition. The stronger $t_3$ response is consistent with coupled changes in active-region overlap, vertical confinement, and resonance alignment. In contrast, $n_1$, $\ell$, and $a$ perturb in-plane feedback and outcoupling together, so improvement in one component can be offset by spectral displacement or weaker extraction.

The action logs connect these local responses to Rainbow-lite's seed dependence. Its strongest concentrated run repeatedly increases $t_3$ and $n_3$, whereas unsuccessful runs devote much of the budget to hole-factor edits and, in one case, decreasing $t_1$. The combined mechanisms can therefore reinforce either a productive or an unproductive direction once an early preference develops, making the observed bimodality physically interpretable without treating the pooled medians as seed-specific causal effects.

Dueling DQN is not associated with one recurring edit: greedy thickness directions differ across seeds, and seed~2 combines policy-guided relocation with stochastic discovery. Panel~(a) further shows that all seven controls are sampled on both sides of their initial values, although over anisotropic ranges. The medians aggregate edges visited at different background geometries rather than a factorial sweep around one cavity; collapsing repeated traversals prevents frequently exploited edges from dominating, but policy selection, correlated trajectories, and missing counterfactuals still preclude a global sensitivity claim. Together with the endpoint and process results, these local responses show that the dueling gain is realized through multiple geometry--response routes rather than one recurring edit or one score component.

\FloatBarrier
\section{Conclusion}

Under a matched 83-call FDTD protocol, Dueling DQN is the only tested modification that improves every seed and produces the strongest endpoint, sample-efficiency, and physical-response metrics. Relative to the first evaluated designs, its selected structures increase the mean $Q$ by 25.8-fold, reduce wavelength error by 64\%, and increase upward power by 47\%, showing that the optimization yields substantial improvements in resonance quality, spectral alignment, and vertical extraction. Distributional, PER, and Noisy DQN alter different parts of the search without a uniform endpoint gain; Double DQN is a controlled null case; and Rainbow-lite exposes both the upside and instability of component interaction. Together, endpoint, process, action-log and physical-response indicate that separating neighborhood value from action advantage is useful when local edits are difficult to rank from sparse coupled feedback. The study therefore contributes both a reproducible comparison framework for PCSEL optimization and a simulation-budget-aware strategy for allocating scarce scientific queries. Beyond the present task, this protocol can also support auditable algorithm selection in other simulation-driven photonic problems. Future work will test its scalability on higher-dimensional PCSEL structures, incorporate fabrication constraints and process tolerances, and validate optimized devices through fabrication and optical measurements.

\section*{Limitations and Ethical Considerations}

This study is limited to simulation-based evaluation; the optimized PCSELs have not yet been fabricated or optically characterized. Re-simulation of the saved geometries verifies numerical consistency, but not fabrication tolerance or measured laser performance. Results are confined to the present PCSEL parameterization and objective, four matched seeds, and an 83-call budget, and do not establish a universal algorithm ranking.

The study uses only algorithm-generated geometries and FDTD responses, without human participants or personal data; privacy, informed-consent, and demographic-bias concerns therefore do not arise. Methodological bias may nevertheless enter through simulator assumptions, design bounds, and objective weights. Potential misuse includes treating simulated rankings as fabrication-ready guarantees; experimental validation is required before claims about fabricated-device performance.

\section*{Generative AI Usage}

OpenAI Codex (GPT-5-based; July 2026) assisted with language. It did not generate FDTD data or results; the authors verified all outputs and retain full responsibility.

\bibliographystyle{ACM-Reference-Format}
\bibliography{references}

\appendix

\section*{Appendix}

For readability, the supplementary material is organized into four sections:
\begin{itemize}
  \item \hyperref[app:endpoint_audit]{Appendix A: Endpoint and Device Audits};
  \item \hyperref[app:environment]{Appendix B: Optimization Environment};
  \item \hyperref[app:implementation]{Appendix C: Algorithm and Training Details}; and
  \item \hyperref[app:process]{Appendix D: Extended Process Diagnostics}.
\end{itemize}

\section{Endpoint and Device Audits}
\label{app:endpoint_audit}

\begin{figure*}[!t]
\centering
\includegraphics[width=0.92\textwidth]{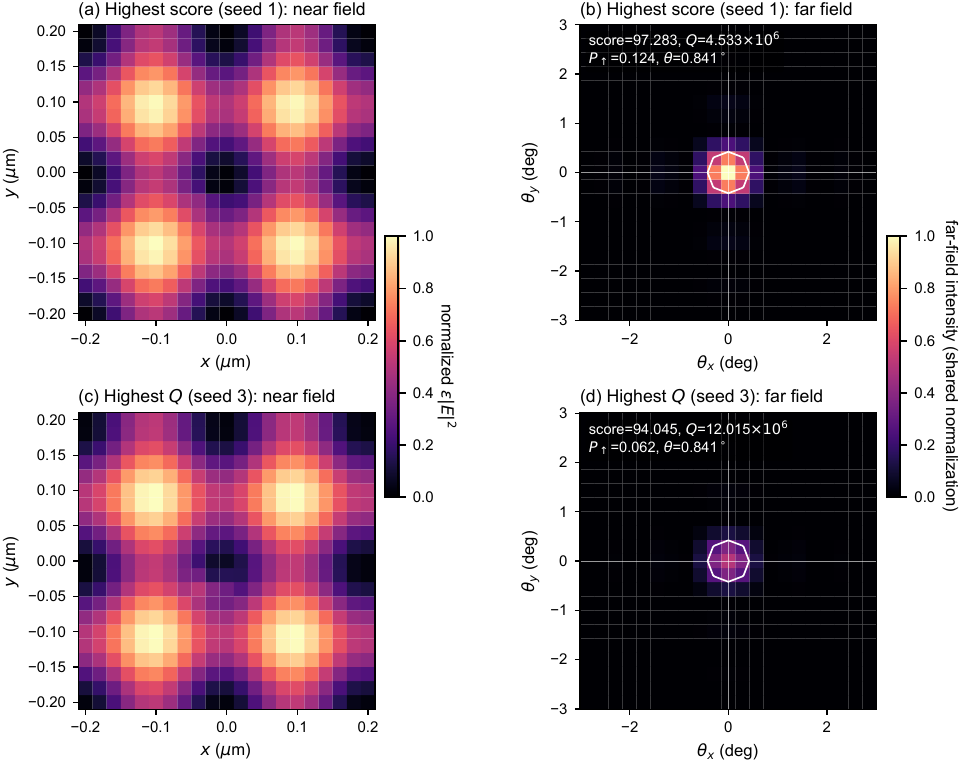}
\Description{Four field maps compare the highest-scoring and highest-Q Dueling DQN endpoints. The left panels show unit-cell-plane near-field energy density, and the right panels show far-field intensity on a common scale. The highest-scoring design has approximately twice the far-field intensity of the highest-Q design.}
\caption{Saved near- and far-field profiles of two representative Dueling DQN endpoints: (a,b) the highest-scoring design (seed 1) and (c,d) the highest-$Q$ design (seed 3). The near-field panels show $\varepsilon|E|^2$ over the central $0.4\times0.4~\mu\mathrm{m}^2$ unit-cell plane at the stored field-monitor wavelength closest to the selected 1310.132-nm resonance (1312.041 nm); each is normalized independently. The far-field panels are evaluated at the closest stored top-monitor wavelength (1310.251 nm) and share a normalization referenced to the peak of the highest-scoring design; white contours mark half of each panel's local maximum. Both designs emit on axis with a divergence of $0.841^\circ$, whereas the highest-scoring design has approximately twice the upward power (0.124 versus 0.062).}
\label{fig:representative_fields}
\end{figure*}

Table~\ref{tab:dqn_variants} retains the per-seed endpoints and first-best calls removed from the main text. Table~\ref{tab:physical_metrics} reports the physical outputs confirmed by re-simulating the best dueling devices, and Figure~\ref{fig:representative_fields} compares the saved fields of the highest-score and highest-$Q$ endpoints.

\begin{table*}[!t]
\centering
\caption{Per-seed best scores under the matched 83-call protocol. Numbers after @ give the first FDTD call attaining that score; gains are relative to matched DQN.}
\label{tab:dqn_variants}
\resizebox{0.98\textwidth}{!}{\begin{tabular}{rrrrrrrrrrrrrr}
\toprule
Seed & DQN best & Dueling best & Dueling gain & Dist. best & Dist. gain & PER best & PER gain & Noisy best & Noisy gain & Double best & Double gain & Rainbow best & Rainbow gain \\
\midrule
0 & 85.20@20 & 88.70@47 & +3.50 & 85.20@20 & +0.00 & 85.20@20 & +0.00 & 85.20@20 & +0.00 & 85.20@20 & +0.00 & 96.54@60 & +11.34 \\
1 & 87.66@67 & 97.28@64 & +9.63 & 88.13@67 & +0.47 & 86.18@51 & -1.48 & 86.14@40 & -1.52 & 87.66@67 & +0.00 & 95.43@35 & +7.78 \\
2 & 92.51@79 & 94.37@42 & +1.86 & 86.33@58 & -6.18 & 84.72@56 & -7.80 & 85.31@54 & -7.21 & 92.51@79 & +0.00 & 78.62@64 & -13.90 \\
3 & 84.65@70 & 94.04@78 & +9.39 & 84.04@54 & -0.61 & 93.88@78 & +9.22 & 94.18@66 & +9.53 & 84.65@70 & +0.00 & 81.46@24 & -3.20 \\
\midrule
Mean & 87.51 & 93.60 & +6.09 & 85.92 & -1.58 & 87.49 & -0.01 & 87.70 & +0.20 & 87.51 & +0.00 & 88.01 & +0.51 \\
\bottomrule
\end{tabular}
}
\end{table*}

\begin{table*}[!t]
\centering
\caption{Best dueling-DQN devices confirmed by re-simulation. $A_\mathrm{eff}$ is in $10^{-13}$~m$^2$.}
\label{tab:physical_metrics}
\resizebox{0.60\textwidth}{!}{\begin{tabular}{rrrrrrrr}
\toprule
Seed & Score & Call & $Q$ ($10^6$) & $\lambda$ (nm) & $P$ & $A_\mathrm{eff}$ ($10^{-13}$) & $\theta$ (deg) \\
\midrule
0 & 88.701 & 47 & 1.378 & 1310.514 & 0.104 & 0.811 & 0.841 \\
1 & 97.283 & 64 & 4.533 & 1310.132 & 0.124 & 0.805 & 0.841 \\
2 & 94.371 & 42 & 4.596 & 1309.750 & 0.066 & 0.879 & 0.841 \\
3 & 94.045 & 78 & 12.015 & 1310.132 & 0.062 & 0.830 & 0.841 \\
\bottomrule
\end{tabular}
}
\end{table*}

The independently normalized near-field maps show similar unit-cell mode profiles but do not compare absolute field amplitudes. On the common far-field scale, the peak and integrated intensities of the highest-scoring design are 2.004 times those of the highest-$Q$ design, consistent with the 2.007-fold ratio of their recorded upward powers. Increasing $Q$ from $4.533\times10^6$ to $1.2015\times10^7$ therefore does not narrow the projected beam and is accompanied by weaker vertical extraction, reducing the score from 97.283 to 94.045. The scalar objective consequently favors the more balanced endpoint rather than the structure with the largest $Q$ alone.

\section{Optimization Environment}
\label{app:environment}

Although the physical search has seven active controls, the implementation retains an eight-dimensional compatibility state containing the fixed width coordinate and the seven controls in Table~\ref{tab:environment_controls}. Here $\Delta w=75$ nm remains fixed. The network keeps 16 output slots, but actions 0 and 1 are excluded from random sampling and masked during greedy and bootstrap selection, leaving 14 selectable edits. Observations are normalized componentwise and clipped to $[-1,1]$; the exact order and scale are given in Table~\ref{tab:shared_training_settings}.

\begin{table*}[!t]
\centering
\caption{PCSEL state coordinates, shared initial geometry, admissible absolute ranges, and discrete local edits.}
\label{tab:environment_controls}
\resizebox{0.76\textwidth}{!}{\begin{tabular}{lrrrr}
\toprule
Control & Initial value & Admissible range & Increment & Actions ($+/-$) \\
\midrule
$w$ (masked) & 2075 nm & 1000--3000 nm & 25 nm & 0/1 \\
$t$ & 455.0 nm & 350--550 nm & 2.5 nm & 2/3 \\
$t_1$ & 45.0 nm & 0--200 nm & 2.5 nm & 4/5 \\
$t_3$ & 287.5 nm & 215--415 nm & 2.5 nm & 6/7 \\
$n_1=n_4$ & 3.2035 & 3.0535--3.3535 & 0.005 & 8/9 \\
$n_3$ & 3.4250 & 3.265--3.565 & 0.005 & 10/11 \\
$\ell$ & 0.5796875 & 0.22--0.82 & 0.005 & 12/13 \\
$a$ & 400.0 nm & 300--500 nm & 2.5 nm & 14/15 \\
\bottomrule
\end{tabular}
}
\par\smallskip
\begin{minipage}{0.92\textwidth}
\footnotesize\emph{Note.} Initial values define the shared reset geometry, whereas bounds are centered on the FDTD-template base values. The $n_1$ action changes $n_1$ and $n_4$ together. The radius $r=\ell a/2$ is derived and has no independent action. Width actions 0 and 1 are retained only for compatibility and are masked in all reported runs; $t_4=5000$ nm is fixed.
\end{minipage}
\end{table*}

Calls 1--20 form the matched random-start trajectory; calls 21--83 form 21 three-edit windows. Each window starts from the highest-scoring geometry observed so far, applies three sequential edits, updates the incumbent immediately upon improvement, and then restores the environment to the incumbent. Windowing controls the search state only: replay contains one-step transitions, with no $n$-step return or sequence bonus, and $r_k=S(\mathbf{s}_{k+1})$.

An out-of-range proposal is simulated before being marked terminal; it receives the $-10$ penalty, has a null next state, and the following episode resets to the nominal design. A solver exception is mapped to $Q=0$, $\lambda=1310$ nm, $P=A_\mathrm{eff}=0$, and $\theta=180^\circ$; failure alone is not terminal. None of these failure or terminal cases occurs in the 28 formal logs. Saved-project logs identify Ansys Lumerical 2024 R1 FDTD Solver 8.31.3633 (Windows 64-bit) with a 16-process $1\times2\times8$ layout.

\section{Algorithm and Training Details}
\label{app:implementation}

The isolated variants modify one identifiable component of DQN, whereas Rainbow-lite combines five of them. All configurations otherwise share the state representation, action mask, simulator budget, optimizer, and update schedule. Tables~\ref{tab:shared_training_settings} and~\ref{tab:variant_settings} separate the common settings from the variant-specific mechanisms.

\begin{table*}[!t]
\centering
\caption{Training settings shared by all 28 matched runs.}
\label{tab:shared_training_settings}
\small
\begin{tabular}{@{}p{0.19\textwidth}p{0.75\textwidth}@{}}
\toprule
Setting & Shared implementation \\
\midrule
State scaling & Divide $[\Delta w,\Delta t,\Delta t_1,\Delta t_3,\Delta n_1,\Delta n_3,\Delta\ell,\Delta a]$ by $[1500,150,150,150,0.225,0.225,0.45,150]$ and clip componentwise to $[-1,1]$; dimensional entries are in nm. \\
Network and actions & ReLU MLP $8$--$80$--$120$--$80$ with 16 compatibility outputs; width actions 0 and 1 are masked, leaving 14 selectable edits. \\
Optimization & Adam, learning rate $10^{-4}$; $\gamma=0.98$; batch size 16; replay capacity 3500. Scalar agents use Smooth-L1 loss and C51 agents use projected categorical cross-entropy. \\
Gradient and target updates & Each gradient element is clipped to $[-1,1]$; one gradient update follows each learned call, and the target network is copied every 50 action selections. \\
Simulation budget & 83 FDTD calls per run. Calls 1--20 are forced-uniform-random transitions stored without learning; learning starts at call 21. \\
Epsilon schedule & $\epsilon_k=0.10+0.55\exp(-k/500)$, where $k$ includes the 20 initialization calls; Noisy DQN and Rainbow-lite set post-initialization epsilon to zero. \\
Local-search protocol & Calls 21--83 form 21 three-edit incumbent-reset windows. Replay remains one-step, $r_k=S(\mathbf{s}_{k+1})$, and the sequence-bonus weight is zero. \\
Reproducibility & Seeds 0--3 set Python, NumPy, and PyTorch generators; all 28 selected configurations record CPU execution. \\
\bottomrule
\end{tabular}

\end{table*}

\begin{table*}[!t]
\centering
\caption{Variant-specific network, target, replay, and exploration settings. Unlisted settings are inherited from Table~\ref{tab:shared_training_settings}.}
\label{tab:variant_settings}
\small
\begin{tabular}{@{}p{0.13\textwidth}p{0.43\textwidth}p{0.16\textwidth}p{0.17\textwidth}@{}}
\toprule
Agent & Output or target modification & Replay & Exploration \\
\midrule
DQN & Scalar 16-action head; standard target-network maximum. & Uniform & Epsilon-greedy \\
Dueling DQN & Scalar value and 16-action advantage heads; advantages are mean-centered over all 16 compatibility outputs before masking. & Uniform & Epsilon-greedy \\
Distributional DQN & $16\times51$ categorical outputs and C51 projection on $[0,5000]$. & Uniform & Epsilon-greedy \\
PER-DQN & Baseline scalar network and target. & Proportional PER & Epsilon-greedy \\
Noisy DQN & Factorized-Gaussian NoisyLinear second and third hidden layers and output head. & Uniform & Parameter noise \\
Double DQN & Online-network action selection and target-network evaluation. & Uniform & Epsilon-greedy \\
Rainbow-lite DQN & Noisy dueling categorical head with a Double target; combines Dueling, C51, PER, NoisyLinear, and Double DQN without multi-step returns. & Proportional PER & Parameter noise \\
\bottomrule
\end{tabular}

\par\smallskip
\begin{minipage}{0.94\textwidth}
\footnotesize\emph{Note.} PER uses proportional priorities with $\alpha=0.6$, $\beta_0=0.4$ annealed to 1 over 1000 action selections, and $\epsilon_\mathrm{PER}=10^{-6}$. NoisyLinear uses factorized Gaussian noise with $\sigma_\mathrm{init}=0.5$. Noisy DQN and Rainbow-lite disable epsilon after initialization.
\end{minipage}
\end{table*}

\section{Extended Process Diagnostics}
\label{app:process}

\begin{figure*}[!t]
\centering
\includegraphics[width=0.82\textwidth]{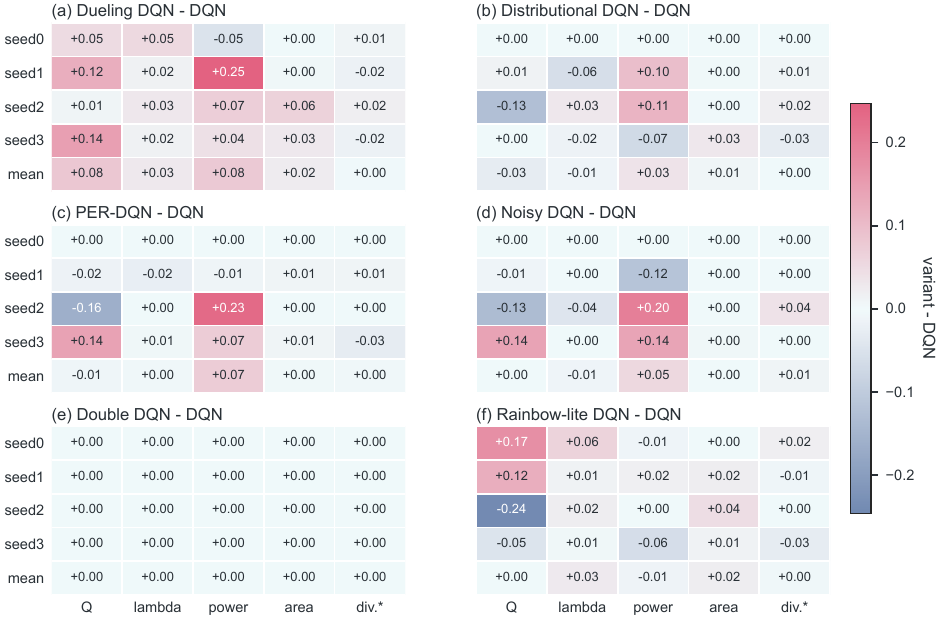}
\Description{Six heatmaps showing seed-level normalized physical-component differences between each matched variant and DQN, together with a cross-seed mean row.}
\caption{Seed-level physical-component differences relative to DQN. Panels (a)--(f) show dueling DQN, distributional DQN, PER-DQN, Noisy DQN, Double DQN, and Rainbow-lite. Each panel reports seeds 0--3 and their mean; positive values indicate larger variant components.}
\label{fig:component_delta_by_seed}
\end{figure*}

\begin{figure*}[!t]
\centering
\includegraphics[width=0.78\textwidth]{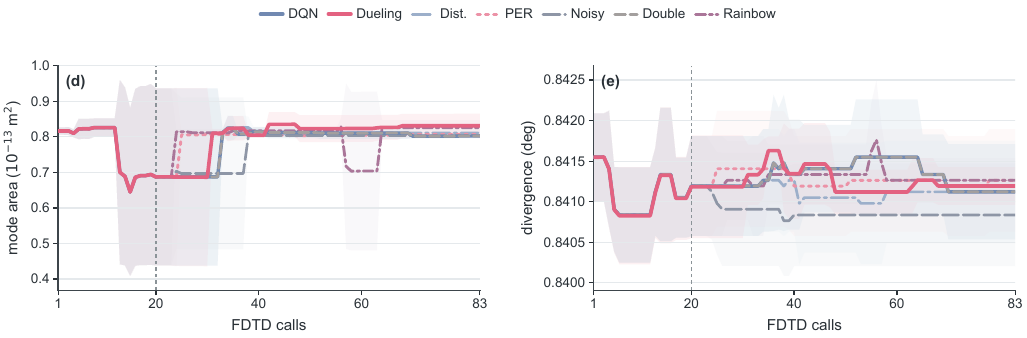}
\Description{Two side-by-side process-level panels comparing seven matched DQN configurations over 83 FDTD calls. Panel d shows effective mode area and panel e shows far-field divergence. Curves are cross-seed means for the incumbent best design, and shaded bands indicate one standard deviation.}
\caption{Supplementary incumbent physical trajectories completing Figure~\ref{fig:dqn_dueling_physical_process}: (d) effective mode area and (e) far-field divergence. Curves show cross-seed means and shaded bands show one standard deviation; the vertical line marks the end of the 20-call initialization period.}
\label{fig:physical_process_supplement}
\end{figure*}

Table~\ref{tab:extended_benchmark_matrix} consolidates the process and multi-metric diagnostics in one comparison. Panel (a) reports endpoint, efficiency, discovery, policy, and physical-response measures; panel (b) adds only the complementary process statistics. Non-exploratory columns denote non-epsilon actions, or noisy-greedy actions for Noisy DQN and Rainbow-lite. Figures~\ref{fig:component_delta_by_seed} and~\ref{fig:physical_process_supplement} provide seed-level component changes and the two secondary incumbent metrics.

\begin{table*}[!t]
\centering
\caption{Extended comparison over four matched seeds. Panel (a) is the comprehensive benchmark; panel (b) adds complementary process statistics without repeating endpoint and threshold columns. Higher is better except for standard deviation, first-threshold call, wavelength error, and average rank. First-threshold calls are conditional on successful seeds.}
\label{tab:extended_benchmark_matrix}
{\small\textit{(a) Comprehensive benchmark.}}\par\smallskip
\resizebox{\textwidth}{!}{%
\begin{tabular}{lrrrrrrrrrrrrrr}
\toprule
& \multicolumn{3}{c}{Endpoint quality} & \multicolumn{2}{c}{Sample efficiency} & \multicolumn{3}{c}{Candidate discovery} & Policy & \multicolumn{4}{c}{Physical response} & Overall \\
\cmidrule(lr){2-4}\cmidrule(lr){5-6}\cmidrule(lr){7-9}\cmidrule(lr){10-10}\cmidrule(lr){11-14}\cmidrule(lr){15-15}
Model & Mean best & Min best & Std. & AUC & First $\geq$83 & $\geq$90 seeds & Top-5 & $\geq$90 calls & $\geq$83 calls & $Q$ ($10^6$) & $|\Delta\lambda|$ & $P$ & $A_\mathrm{eff}$ ($10^{-13}$) & Avg. rank \\
\midrule
DQN & 87.51 & 84.65 & 3.59 & 84.35 & 37.2 & 1/4 & 85.77 & 0.50 & 3.00 & 2.19 & 0.86 & 0.078 & 0.80 & 4.54 \\
\textbf{Dueling} & \textbf{93.60} & \textbf{88.70} & 3.58 & \textbf{89.22} & 29.8 & \textbf{3/4} & \textbf{93.12} & \textbf{12.50} & 7.25 & \textbf{5.63} & \textbf{0.26} & \textbf{0.089} & \textbf{0.83} & \textbf{1.23} \\
Dist. & 85.92 & 84.04 & \textbf{1.74} & 83.76 & 37.8 & 0/4 & 84.92 & 0.00 & 5.00 & 1.32 & 0.95 & 0.083 & 0.81 & 5.54 \\
PER & 87.49 & 84.72 & 4.30 & 84.08 & 40.5 & 1/4 & 85.25 & 0.50 & 1.00 & 2.61 & 0.93 & 0.089 & 0.81 & 4.65 \\
Noisy & 87.70 & 85.20 & 4.34 & 85.52 & 32.5 & 1/4 & 85.16 & 0.25 & 5.50 & 1.83 & 0.98 & 0.086 & 0.80 & 4.35 \\
Double & 87.51 & 84.65 & 3.59 & 84.35 & 37.2 & 1/4 & 85.77 & 0.50 & 3.00 & 2.19 & 0.86 & 0.078 & 0.80 & 4.54 \\
Rainbow & 88.01 & 78.62 & 9.29 & 84.78 & \textbf{21.5} & 2/4 & 87.22 & 11.00 & \textbf{15.75} & 2.40 & 0.35 & 0.075 & 0.83 & 3.15 \\
\bottomrule
\end{tabular}

}
\par\medskip
{\small\textit{(b) Complementary process statistics.}}\par\smallskip
\resizebox{0.68\textwidth}{!}{%
\begin{tabular}{lrrrr}
\toprule
Model & Post-init. mean & Calls 64--83 mean & $\geq$83 calls & Non-exploratory mean \\
\midrule
DQN & 74.88 & 76.79 & 8.75 & 73.55 \\
Dueling & 77.40 & 79.69 & 21.00 & 75.34 \\
Dist. & 74.66 & 76.34 & 9.00 & 73.84 \\
PER & 74.26 & 75.50 & 6.50 & 73.44 \\
Noisy & 74.54 & 74.98 & 5.50 & 74.54 \\
Double & 74.88 & 76.79 & 8.75 & 73.55 \\
Rainbow & 77.67 & 80.40 & 15.75 & 77.67 \\
\bottomrule
\end{tabular}

}

\end{table*}

\FloatBarrier

\end{document}